\documentclass[11pt]{article}
\usepackage{jheppub}
\pdfoutput=1
\usepackage{amsmath}
\usepackage{amssymb}
\usepackage{graphics}
\usepackage{pdfsync}
\usepackage{shuffle}
\usepackage{subfigure}
\usepackage{mathtools}
\usepackage[table]{xcolor}
\usepackage{hyperref}
\hypersetup{colorlinks=true,linkcolor=magenta,anchorcolor=green,citecolor=cyan,filecolor=black,menucolor=black,urlcolor=brown}
\preprint{IPHT-t18/003}
\title{A Vertex Operator Algebra Construction of the Colour-Kinematics Dual numerator}
\author[a]{Chih-Hao Fu,}
\author[b,c]{Pierre Vanhove}
\author[d]{\textrm{and} Yihong Wang}

\affiliation[a]{School of Physics and Information Technology,
Shaanxi Normal University,\\
No.620 West Chang'an Avenue, Xi'an 710119, P.R. China.}
\affiliation[b]{Institut de Physique Th\'eorique,\\
  Universit\'e Paris Saclay, CNRS, F-91191 Gif-sur-Yvette, France}
\affiliation[c]{National Research University Higher School of Economics, Russian Federation}
\affiliation[d]{Department of Physics, National Taiwan University,\\
No.1 Sec.4 Roosevelt Road Taipei 10617,Taiwan (R.O.C.).}

\emailAdd{chihhaofu@snnu.edu.cn}
\emailAdd{pierre.vanhove@ipht.fr}
\emailAdd{yihongwang@phys.ntu.edu.tw}

\date{\today} \abstract{We derive a vertex operator based expression
  for the kinematic numerators of Yang-Mills amplitudes by applying
  the momentum kernel formalism to open string amplitudes.  The
  expression involves an $\alpha'$-weighted commutator induced by the
  monodromy relations between the colour ordered Yang-Mills
  amplitudes, which mirrors the $\alpha'$ deformed colour structure
  observed in open string and semi-abelian $Z$-theory.  The kinematic
  algebra given by this construction contains the Lie algebra of
  diffeomorphism as an obvious sub-algebra.

}

\begin{document}
\maketitle \flushbottom

\section{Introduction}
\label{SecIntroduction}
The discovery of colour-kinematics duality~\cite{Bern:2008qj,Bern:2010ue} has led to numerous
insights in the nature of gauge theory, gravity theory and their
relations.  Assuming a cubic graph representation
of a Feynman diagram the colour-kinematics duality
states that the kinematic numerators satisfy the same algebraic
relations as the colour factor associated with the same diagram. At
tree-level this duality implies kinematic relations between colour
ordered amplitudes in gauge theory~\cite{Bern:2008qj} which have been
proven for tree-level graphs using string theory
methods~\cite{BjerrumBohr:2009rd,Stieberger:2009hq} or quantum field
theory
techniques~\cite{Feng:2010my,BjerrumBohr:2010ta,BjerrumBohr:2010yc,Chen:2011jxa}. This
duality provides a simple and powerful rule for constructing gravity amplitudes
from gauge theory amplitudes by replacing colour factors with dual
kinematics factors~\cite{Bern:2010ue,Bern:2010yg}. At tree-level the
double copy procedure has been proven~\cite{BjerrumBohr:2010ta,BjerrumBohr:2010yc}, and shown to be equivalent to the
Kawai-Lewellyn-Tye (KLT) relations~\cite{Kawai:1985xq} between closed
string and open string amplitudes~\cite{BjerrumBohr:2010hn}. This duality has
been used in constructing tree amplitudes~\cite{Anastasiou:2016csv,Anastasiou:2017nsz,Johansson:2017srf,Chiodaroli:2017ehv,Johansson:2018ues,Chiodaroli:2015wal} and loop amplitudes in various
supergravity
theories~\cite{Bern:2011rj,BoucherVeronneau:2011qv,Bern:2012uf,Bern:2012cd,Bern:2012gh,Carrasco:2012ca,Bern:2013yya,Bern:2013qca,Bern:2013uka,Bern:2014lha,Bern:2014sna,
  Johansson:2014zca,Chiodaroli:2014xia}, classical general relativity solutions~\cite{Goldberger:2017frp,Luna:2016hge,Luna:2016due,Ridgway:2015fdl,Luna:2015paa,Monteiro:2014cda,Adamo:2017nia,Carrillo-Gonzalez:2017iyj,Goldberger:2017ogt,Li:2018qap}, and a host of quantum field
theories~\cite{Broedel:2012rc,Chiodaroli:2013upa,Nohle:2013bfa,Carrasco:2015iwa,Chiodaroli:2017ngp,Chen:2013fya,Cachazo:2014xea,Carrasco:2016ldy}. 
At five-loop order a generalised double copy method allowed to pin
down the critical ultraviolet behaviour of the four-graviton amplitude in
maximal supergravity~\cite{Bern:2017ucb,Bern:2018jmv}. Such generalised
Jacobi-like relations arise from the most general solution of the monodromy
relations between the colour ordered gauge theory
amplitudes~\cite{BjerrumBohr:2010zs}.
Assuming the validity of the colour-kinematics duality to all loop orders one can derive the
critical ultraviolet behaviour of the four-graviton amplitude in
maximal supergravity~\cite{Vanhove:2010nf}.
 This strengthens the idea of an
underlying principle responsible for the colour-kinematics duality. Another piece of evidence supporting this idea is that
the Lie algebra of diffeomorphism has been identified to give rise to
the kinematic numerators at least for special helicity configurations
and for small multiplicity
in~\cite{Monteiro:2011pc,BjerrumBohr:2012mg,Fu:2012uy,Fu:2016plh}.

A clue that may help to solve our above puzzle comes from a property
of CFT known to the quantum groups community. It is known that a CFT,
can be used to build not
only a Lie algebra, but a much richer-in-structure Hopf algebra \cite{fuchs-quantumgroups}.
Indeed, as a matter of fact a similar argument was used in \cite{Green:1987sp}
to build the global $E_{8}\times E_{8}$ symmetry generators of the
heterotic string when kinematic restrictions were imposed due to the
compactification condition. Considering the relation to heterotic
string theory and to its $\mathcal{N}=4$ supersymmetric version at
tree level, we feel it is reasonable to suspect that certain weaker
version of the Hopf algebraic structure survives in Yang-Mills. 
 A hint that may be related to this structure was recently observed in 
\cite{Fu:2016plh}, where it was demonstrated that the Yang-Mills 
cubic vertex can be obtained  as a projected bracket of the Drinfeld
double constructed naturally by regarding gauge fields as vector
fields supplemented with dual one-forms. The projection broke the Jacobi identity 
which  was shown to be restored once the quartic vertex contribution is included.

In this paper we carry the spirit discussed above one step further
and investigate the Jacobi-like relations between Yang-Mills kinematic
numerators from $\alpha'$ limit of the open string amplitudes. We
show at least from string perspective there is genuinely a natural
cubic graph description of the scattering amplitude derived from the
vertex operator algebra that when reaching the $\alpha'\rightarrow0$
limit satisfies the anti-symmetry and Jacobi identities assumed by
the BCJ duality. In particular, we find a half-ladder basis numerator
is given by the following simple expression 
\begin{equation}
n(123\dots n)=\lim_{k_{n}^{2},\alpha'\rightarrow0}\Bigl\langle 
f\Bigr|[[[T_{1},T_{2}]_{\alpha'},T_{3}]_{\alpha'}\dots,T_{n-1}]_{\alpha'}\Bigl|
0\Bigr\rangle,\label{eq:result-expression}
\end{equation}
where $|f\rangle$ is the modified external
  state defined in~\eqref{e:modified}  and  the generators $T_{i}$ here are vertex operators, being properly
analytic continued and integrated, $T_{i}=\int_{0}^{1}dz\,V_{i}(z)/z$,
and $[\;,\;]_{\alpha'}$ is the $\alpha'$-weighted commutator
\begin{equation}
[T_{1},T_{2}]_{\alpha'}=T_{1}T_{2}-e^{-i\pi\alpha'k_{2}\cdot k_{1}}T_{2}T_{1}
\end{equation}
 originally
introduced in \cite{Ma:2011um} to express the string theory generalization
of the Del Duca-Dixon-Maltoni \cite{DelDuca:1999rs} colour decomposition of Yang-Mills
amplitude 
\begin{equation}\label{e:DDM}
\mathcal{M}^{\rm open}_{n}=\sum_{\sigma\in S_{n-2}}tr([[[t_{1},t_{\sigma(2)}]_{\alpha'},t_{\sigma(3)}]_{\alpha'}\dots,t_{\sigma(n-1)}]_{\alpha'}t_{n})\,\mathcal{A}_{n}(1,\sigma(2),\dots,\sigma(n-1),n)
\end{equation}
derived from string monodromy relation, and recently again observed
in semi-abelian $Z$-theory in~\cite{Carrasco:2016ygv}.  The construction of the
present 
paper gives a  kinematic analogue of the colour traces in~\eqref{e:DDM}
and provides an alternative construction to the kinematic traces
of~\cite{Bern:2011ia}. 
The construction
of this paper  uses bosonic string theory, but the discussion
generalises easily to the superstring case. In fact  similar
structures have been obtained using conformal blocks in~\cite{Mafra:2011kj} using the pure
spinor formalism in   open string theory, and have been recently generalised to heterotic and type II strings,
in particular to one loop level
in \cite{Ochirov:2013xba}.  
From a string perspective this answers the question raised earlier above
as to why the product of momentum kernel with colour-ordered amplitude
should possess algebraic properties. In addition, we identify the Lie
algebra of diffeomorphism as the vector $\times$ vector $\rightarrow$
vector part of the sub-algebra, while the full numerator is restored
when scalar and tensor contributions are included. The kinematic algebra
obtained as the field theory limit of vertex operator algebra in this
paper however has the apparent drawback of being lack of simplicity.
At the moment it is not completely clear to us whether a wiser representation
exists or the algebraic expression is of any practical use. We hope
that perhaps its analytic feature can be taken as a useful reference
when constructing numerator ansatz at higher loop orders. The vertex
algebra based numerators (\ref{eq:result-expression}) derived in
this paper demonstrate another formal symmetry between the colour
and kinematic factors of the string amplitude, in an expression that
is even closer to the field theory double copy structure.

This paper is organised as follows. In section \ref{sec:review} we
briefly review a few analytic features of the string KLT monodromy relations,
especially the monodromy related properties presented in~\cite{BjerrumBohr:2010hn},
which will prove very much useful in our later discussions. We next
introduce in section \ref{sec:off-shell-current} a specific off-shell
continuation of the open string amplitude that will serve our purpose
of deriving a vertex operator explanation. The relations between vertex
operator algebra and BCJ numerators will be unravelled through two
examples in sections \ref{sec:the-algebra-3pt} and \ref{sec:the-algebra-4pt},
followed by a short discussion on the explicit form of the generators
that appears in the numerator formula. In section \ref{sec:the-diffeo}
we reproduce the Lie algebra of diffeomorphism as a sub-algebra. In
section \ref{sec:hyper-geometric-functions} we will demonstrate generically
what analytic structure appears in a string of structure constants
at higher points and its relation to hypergeometric functions arising
from disc integrals. We conclude
the paper with a brief comment on our results and related problems
in section \ref{sec:conclusion}.

\section{Preliminaries}
\label{sec:review}

\begin{figure}[t]
\centering
\subfigure[]{
\includegraphics[width=8cm]{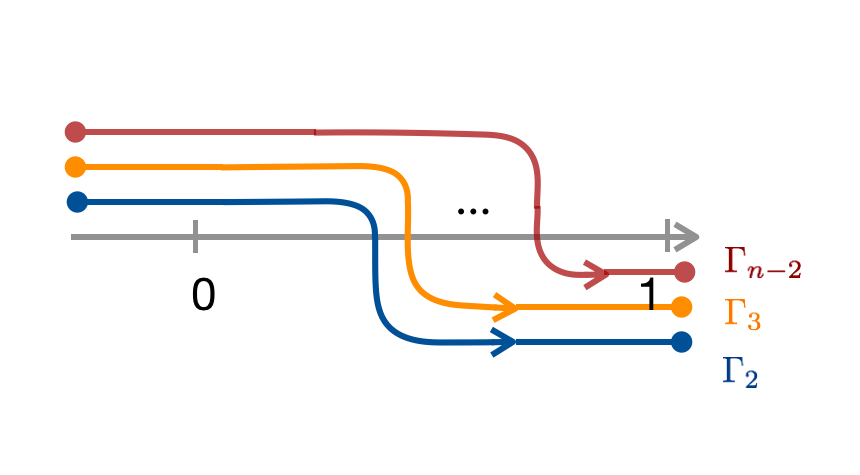}
 \label{fig:contours-1}}
\subfigure[]{
\includegraphics[width=8cm]{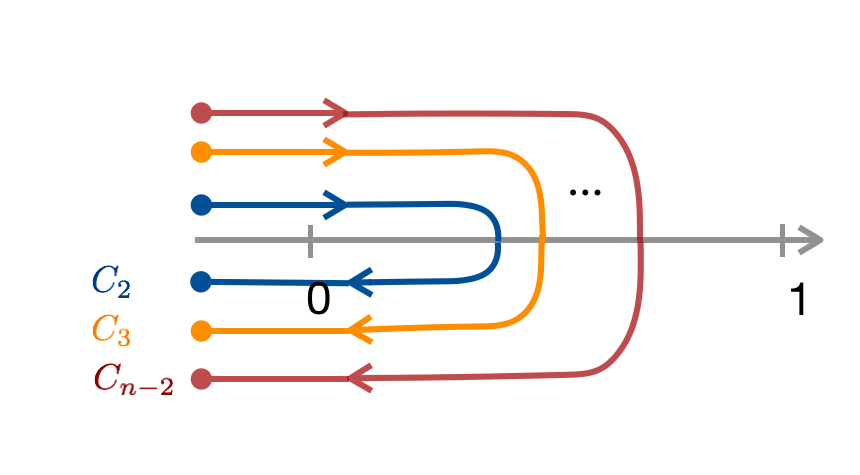}
\label{fig:contours-2}}
\caption{Integration contours associated with world-sheet variables $v_{i}^{-}$
in the KLT monodromy relations.}
\end{figure}
In this section we review a few details related to the string KLT monodromy
relations that will become useful to our later derivations. It was
demonstrated in \cite{BjerrumBohr:2010hn} that when properly analytic continued,
the world-sheet integral of a bosonic closed string amplitude factorises
according to its dependence on light-cone coordinates $v_{i}^{\pm}=v_{i}^{1}\pm v_{i}^{2}$,
\begin{equation}
\mathcal{M}^{\rm closed}_{n}=\sum_{\alpha\in S_{n-3}}\mathcal{A}_{n}(1,\alpha,n-1,n)\,\mathcal{I}_{\alpha},\label{eq:klt-string}
\end{equation}
where the $v^{+}$ integrals together was identified as a colour-ordered
bosonic open string amplitude $\mathcal{A}_{n}$, and the $v^{-}$
integrals together reads
\begin{equation}
\mathcal{I}_{23\dots n-2}=\int_{\Gamma_{2}}dv_{2}^{-}\int_{\Gamma_{3}}dv_{3}^{-}\dots\int_{\Gamma_{n-1}}dv_{n-1}^{-}\,\prod_{i=2}^{n-2}(v_{i}^{-})^{\alpha'k_{1}\cdot k_{i}}(1-v_{i}^{-})^{\alpha'k_{n-1}\cdot k_{i}}\prod_{i<j\leq n-2}(v_{i}^{-}-v_{j}^{-})^{\alpha'k_{i}\cdot k_{j}}f(v_i^{-}),
\end{equation}
where $f(v_i^-)$ arises from the operator product expansion of the
vertex operators.
The integration contours $\Gamma_{2},\Gamma_{3},\dots,\Gamma_{n-2}$
illustrated in Fig \ref{fig:contours-1} were defined to avoid branch
cuts due to factors $(v_{i}^{-}-v_{j}^{-})^{\alpha'k_{i}\cdot k_{j}}$,
while variables $(v_{1}^{-},v_{n-1}^{-},v_{n}^{-})$ were gauge fixed
by world-sheet Moebius invariance to be $(0,1,\infty)$ respectively.
These contours were then pulled to the left, assuming no pole lying
in the lower half plane. The result was to replace $\Gamma_{2}$, $\dots$,
$\Gamma_{n-2}$ by a new set of contours defined along the branch
cut $[0,-\infty)$ (Fig \ref{fig:contours-2}). It was noted that
the phase difference between the two sides of the branch cut $[0,-\infty)$
in the $C_{2}$ integration produces an overall sine factor,
\begin{align}
 & \int_{C_{2}}dv_{2}^{-}(v_{2})^{\alpha'k_{1}\cdot k_{2}}(1-v_{2}^{-})^{\alpha'k_{n-1}\cdot k_{2}}\prod_{j+3}^{n-2}(v_{j}^{-}-v_{2}^{-})^{\alpha'k_{j}\cdot k_{2}}f(v_{2}^{-})\nonumber \\
 & =2i\,\sin(\pi\alpha'k_{1}\cdot k_{2})\int_{-\infty}^{0}dv_{2}^{-}(-v_{2}^{-})^{\alpha'k_{1}\cdot k_{2}}(1-v_{2}^{-})^{\alpha'k_{n-1}\cdot k_{2}}\prod_{j+3}^{n-2}(v_{j}^{-}-v_{2}^{-})^{\alpha'k_{j}\cdot k_{2}}f(v_{2}^{-}).\label{eq:v2-integral}
\end{align}
and similarly for the contour $C_{3}$
\begin{align}
 & \int_{C_{3}}dv_{3}^{-}(v_{3}^{-})^{\alpha'k_{1}\cdot k_{3}}(1-v_{3}^{-})^{\alpha'k_{n-1}\cdot k_{3}}(v_{3}^{-}-v_{2}^{-})^{\alpha'k_{2}\cdot k_{3}}\prod_{j=4}^{n-2}(v_{j}^{-}-v_{3}^{-})^{\alpha'k_{j}\cdot k_{3}}f(v_{3}^{-})\nonumber \\
 & =2i\sin(\pi\alpha'k_{1}\cdot k_{3})\int_{v_{2}^{-}}^{0}dv_{3}^{-}(-v_{3}^{-})^{\alpha'k_{1}\cdot k_{3}}(1-v_{3}^{-})^{\alpha'k_{n-1}\cdot k_{3}}(v_{3}^{-}-v_{2}^{-})^{\alpha'k_{2}\cdot k_{3}}\prod_{j=4}^{n-2}(v_{j}^{-}-v_{3}^{-})^{\alpha'k_{j}\cdot k_{3}}f(v_{3}^{-})\nonumber \\
 & +2i\sin(\pi\alpha'(k_{1}+k_{3})\cdot k_{3})\int_{-\infty}^{v_{2}^{-}}(-v_{3}^{-})^{\alpha'k_{1}\cdot k_{3}}(1-v_{3}^{-})^{\alpha'k_{n-1}\cdot k_{3}}(v_{2}^{-}-v_{3}^{-})^{\alpha'k_{2}\cdot k_{3}}\prod_{j=4}^{n-2}(v_{j}^{-}-v_{3}^{-})^{\alpha'k_{j}\cdot k_{3}}f(v_{3}^{-})\,.\label{eq:v3-integral}
\end{align}
Repeating the same manipulation on all contours led to the extraction
of a momentum kernel, and the remaining factor was identified as another
copy of the bosonic open string amplitude $\tilde{A}_{n}$,
\begin{equation}
\mathcal{I}_{23\dots n-1}=\sum_{\beta\in S_{n-3}}\mathcal{S}_{\alpha'}[n-1,\dots,4,3,2|\beta]\,\tilde{\mathcal{A}}_{n}(n-1,n,\beta,1),\label{eq:numerator-ampl-eqn}
\end{equation}
where the momentum kernel $\mathcal{S}_{\alpha'}[i_{2},\dots,i_{k}|j_{2},\dots,j_{k}]_{p}$
in the context of string theory is defined as\footnote{ Thanks to the
  factorisation property~\cite[eq~(3.2)]{BjerrumBohr:2010hn} the momentum kernel
  enjoys  a nice
  recursive definition~\cite[eq.(2.7)]{Carrasco:2016ldy}.}
\begin{equation}
\mathcal{S}_{\alpha'}[i_{1},\dots,i_{k}|j_{1},\dots,j_{k}]_{p}:=(\pi\alpha'/2)^{-k}\prod_{t=1}^{k}\sin(\pi\alpha'(p\cdot k_{i_{t}}+\sum_{q>t}^{k}\theta(i_{t},i_{q})k_{i_{t}}\cdot k_{i_{q}})),\label{eq:string-momentum-kernel}
\end{equation}
where $\theta(i_t,i_q)=1$ if the ordering of $i_t$ and $i_q$ is the
opposite in the $\{i_1,\dots,i_k\}$ and $\{j_1,\dots,j_k\}$ and
otherwise 0 if the ordering is the same.
This expression  becomes identical to the field theory momentum kernel~\cite{BjerrumBohr:2010ta} in the
$\alpha'\rightarrow0$ limit
\begin{equation}
\mathcal{S}[i_{1},\dots,i_{k}|j_{1},\dots,j_{k}]_{p}=\prod_{t=1}^{k}(p\cdot k_{i_{t}}+\sum_{q>t}^{k}\theta(i_{t},i_{q})k_{i_{t}}\cdot k_{i_{q}}).\label{eq:qft-momentum-kernel}
\end{equation}

In the literature of colour-kinematics duality, it was realised that
the momentum kernel relations described above offer a convenient solution to the
kinematic numerators~\cite{BjerrumBohr:2010zs,Bern:2010yg}. This is because if we treat
the gravity amplitude as double copies and leave one copy of the Yang-Mills
amplitude as it is, the other copy combines with the momentum kernel,
which is understood to be the inverse of the propagator matrix, and
produces an $(n-2)!$-basis half ladder numerator $n(1\alpha n)$
associated with that copy of the amplitude. The result is the familiar
Del Duca-Dixon-Maltoni expression,
\begin{equation}
\mathcal{M}^{\rm closed}_{n}=\sum_{\alpha\in S_{n-2}}\mathcal{A}_{n}(1,\alpha,n)\,n(1\alpha n).\label{eq:ddm-eqn}
\end{equation}
Comparing equation (\ref{eq:ddm-eqn}) with its string theory analogue
(\ref{eq:klt-string}), it is natural to conclude that the $v^{-}$
integral $\mathcal{I}_{\alpha}$ becomes the basis numerator $n(1\alpha n)$
in the $\alpha'$ limit.  One can express the half ladder numerator
using the momentum kernel~\cite{BjerrumBohr:2010hn}  as 
\begin{equation}
n(1,\gamma(2),\gamma(3),\dots,\gamma(n-1),n)=\begin{cases}
\sum_{\beta\in S_{n-3}}\mathcal{S}[\gamma^{T}|\beta]\,\tilde{\mathcal{A}}_{n}(1,\beta,n,n-1) & ,\gamma(n-1)=n-1\\
0 & ,\gamma(n-1)\neq n-1\,.
\end{cases}\label{eq: n-3-prescription}
\end{equation}
These numerators are not unique as any shifts proportional to the 
momentum kernel will not change the total amplitude~\eqref{eq:ddm-eqn}. 
These numerators are  not always local and can present
poles~\cite{Bern:2010yg} but the total amplitude is always local.
These non-localities are sometime useful in finding a colour-kinematics
representation of gauge theory amplitudes~\cite{Carrasco:2011mn},
which is consistent with (\ref{eq:numerator-ampl-eqn}). The above
prescription however has the apparent shortcoming that not all of
its legs are treated on equal footing (which could be regarded as
the result of a generalised gauge shift on the numerators) making
it difficult to allow algebraic interpretation. It is also known that
when applied to amplitudes that has an algebraic structure by construction
such as those of the $\phi^{3}$ theory, an $(n-3)!$ minimal KLT basis
(\ref{eq: n-3-prescription}) yields shifted numerators rather than
the expected string of structure constant $f^{ab*}f^{*c*}f^{*d*}\dots$
\footnote{This can be easily verified, for example by a straightforward calculation
at four points.}. In view of these a slight modification to this approach was added
in \cite{Mafra:2016ltu,Du:2016tbc} where the numerators were solved in a more
symmetric $(n-2)!$ basis
\begin{equation}
n(1,\gamma(2),\gamma(3),\dots,\gamma(n-1),n)=\lim_{k_{n}^{2}\rightarrow0}\sum_{\beta\in S_{n-2}}\frac{1}{k_{n}^{2}}\mathcal{S}[\gamma^{T}|\beta]\,\tilde{\mathcal{A}}_{n}(1,\beta,n)\,,\label{eq:n-2-qft-numerator}
\end{equation}
at the cost that one of the legs must be taken off-shell until the
end of the calculation in order to keep a $(n-2)!$ basis momentum
kernel matrix non-singular.

In the following sections we derive the kinematic algebra from string
perspective using similar reasoning to that described above, except
backwards. Starting with an $(n-2)!$ basis prescription for kinematic
numerator (\ref{eq:n-2-qft-numerator}), we translate the factors
of sine introduced by momentum kernel as world-sheet integrals along
two sides of a branch cut. (In light of the fact that an $(n-2)!$ basis
prescription does lead to unshifted numerator for $\phi^{3}$ theory.)
For this purpose an off-shell continuation to the string amplitude
is introduced. As we shall see, the numerator thus defined does have
a natural algebraic explanation. 

\section{An off-shell continuation of the open string amplitude}
\label{sec:off-shell-current}

For the purpose of discussion we recall that an $n$-point bosonic
open string amplitude is defined in the operator language as
\begin{equation}
\mathcal{A}_{n}=\alpha'^{n-3}g^{n-2}\Bigl\langle f\Bigr|V_{n-1}(1)\frac{1}{L_{0}-I}V_{n-2}(1)\dots V_{3}(1)\frac{1}{L_{0}-I}V_{2}(1)\Bigl|i\Bigr\rangle\,,\label{eq:open-string-ampl}
\end{equation}
  where the external states are $|f\rangle=\lim_{z_n\to\infty}
   z_nV(z_n) |0\rangle$ and $|i\rangle=\lim_{z_1\to0} z_1^{-1}V(z_1) |0\rangle$ acting on the vacuum
   $|0\rangle$.
In standard calculation \cite{Green:1987sp} the propagators are replaced
by integrals $\frac{1}{L_{0}-I}=\int_{0}^{1}dz\,z^{L_{0}-2}$, yielding
\begin{align}
\mathcal A_n&={\alpha'}^{n-3}\, g^{n-2}\,
  \int_{0}^{1}dz_{n-2}\dots\int_{0}^{1}dz_{3}\int_{0}^{1}dz_{2}\Bigl\langle f\Bigr|V_{n-1}(1)z_{n-2}^{L_{0}-2}V_{n-2}(1)z_{n-3}^{L_{0}-2}V_{n-3}(1)\dots z_{3}^{L_{0}-2}V_{3}(1)z_{2}^{L_{0}-2}V_{2}(1)\Bigl|i\Bigr\rangle\nonumber \\
 & ={\alpha'}^{n-3}\,
   g^{n-2}\,\int_{0}^{1}\frac{dz_{n-2}}{z_{n-2}}\dots\int_{0}^{1}\frac{dz_{3}}{z_{3}}\int_{0}^{1}\frac{dz_{2}}{z_{2}}\Bigl\langle
   f\Bigr|V_{n-1}(1)\Bigl(z_{n-2}^{L_{0}-1}V_{n-2}(1)z_{n-2}^{-(L_{0}-1)}\Bigr)\nonumber \\
&\times\Bigl((z_{n-2}z_{n-3})^{L_{0}-1}V_{n-1}(1)(z_{n-2}z_{n-3})^{-(L_{0}-1)}\Bigr)
 \dots\Bigl((z_{3}z_{4}\dots
  z_{n-2})^{L_{0}-1}V_{3}(1)(z_{3}z_{4}\dots
  z_{n-2})^{-(L_{0}-1)}\Bigr) \nonumber \\
&\times\Bigl((z_{2}z_{3}\dots z_{n-2})^{L_{0}-1}V_{2}(1)(z_{2}z_{3}\dots z_{n-2})^{-(L_{0}-1)}\Bigr)\Bigl|i\Bigr\rangle\,,
\end{align}
where in the second line above we simultaneously multiplied and divided
by factors of $z_{i}^{L_{0}-1}$ between vertex operators. The familiar
Veneziano type world-sheet integral formula is then obtained via a
change of integration variables to 
\begin{align}
y_{2} & := z_{2}z_{3}z_{4}\dots z_{n-2}\,\nonumber \\
y_{3} & := z_{3}z_{4}\dots z_{n-2}\,\nonumber \\
 & \vdots\nonumber \\
y_{n-2} & := z_{n-2}\,
\end{align}
and using the fact that vertex operators are primary fields of conformal
dimension one, $V(y)=y^{L_{0}-1}V(1)y^{-(L_{0}-1)}$,
\begin{align}
\mathcal{A}_{n}(123\dots n) &=\alpha'^{n-3}g^{n-2}\int_{0<y_{2}<y_{3}<\dots<y_{n-2}<1}\prod_{i=2}^{n-2}dy_{i}\,\Bigl\langle f\Bigr|V_{n-1}(1)\,\frac{V_{n-2}(y_{n-2})}{y_{n-2}}\dots\frac{V_{3}(y_{3})}{y_{3}}\,\frac{V_{2}(y_{2})}{y_{2}}\Bigl|i\Bigr\rangle\\
\nonumber & =\alpha'^{n-3}g^{n-2}\int_{0<y_{2}<y_{3}<\dots<y_{n-2}<1}\prod_{i=2}^{n-2}dy_{i}\,y_{2}^{\alpha'k_{2}\cdot k_{1}}(1-y_{2})^{\alpha'k_{2}\cdot k_{n-1}}\prod_{i<j<n-2}(y_{j}-y_{i})^{\alpha'k_{j}\cdot k_{i}}f(y_i),
\end{align}
where $f(y_i)$ arises from the operator product expansion of the
vertex operators.
The integration domain changes as a result to $0<y_{2}<y_{3}<\dots<y_{n-2}<1$.

Instead of equation (\ref{eq:open-string-ampl}), we consider a similar
formula ended however with an off-shell final state,

\begin{align}
\mathcal{J}_{n}(123\dots n) & :=\alpha'^{n-3}g^{n-2}\frac{1}{\hat{k}_{n}^{2}}\Bigl\langle\hat{f}\Bigr|V_{n-1}(1)\frac{1}{L_{0}-I}V_{n-2}(1)\dots\frac{1}{L_{0}-I}V_{2}(1)\Bigl|\hat{i}\Bigr\rangle\nonumber \\
 & =\alpha'^{n-2}g^{n-2}\Bigl\langle\hat{f}\Bigr|\frac{1}{L_{0}-I}V_{n-1}(1)\frac{1}{L_{0}-I}V_{n-2}(1)\dots\frac{1}{L_{0}-I}V_{2}(1)\Bigl|\hat{i}\Bigr\rangle.\label{eq:offshell-continuation}
\end{align}
where the off-shell continuation is defined through BCFW-like shifting
$\hat{k}_{n}=k_{n}+\epsilon q$ and $\hat{k}_{1}=k_{1}-\epsilon q$,
with the shifting chosen such that $q\cdot k_{1}=0$ while $q\cdot k_{n}\neq0$,
and  the initial and final states are defined as
\begin{align}\label{e:modified}
\Bigl|\hat{i}\Bigr\rangle&:=\epsilon_{1}\cdot\alpha_{-1}e^{i(k_{1}-\epsilon q)\cdot x}\Bigl|0\Bigr\rangle\,\\
\nonumber\Bigl|\hat{f}\Bigr\rangle&:=\epsilon_{n}\cdot\alpha_{-1}e^{i(k_{n}+\epsilon q)\cdot x}\Bigl|0\Bigr\rangle\,.
\end{align}
The same off-shell continuation was introduced earlier in field theory
amplitudes in \cite{Feng:2010hd} to define KLT monodromy relations in
an $(n-2)!$
basis. Note that this is different from how the off-shell continuation
would usually be defined on string amplitudes \cite{offshell-string1,Liccardo:1999hk},
since conformal symmetry is respected by vertex operators only when
on-shell, and it makes little sense to define a vertex operator if
we are not allowed to shrink the external leg to a point via conformal
transformation in the first place. Nevertheless, it is apparent that
apart from an overall $1/k_{n}^{2}$, equation (\ref{eq:offshell-continuation})
formally agrees with the amplitude (\ref{eq:open-string-ampl}) in
the on-shell limit. In this sense equation (\ref{eq:offshell-continuation})
is a string analogue of the off-shell current. As in the standard
amplitude calculation discussed above we then replace all propagators,
including the newly introduced one by the off-shell leg, and write
\begin{multline}
\mathcal{J}_{n}(123\dots n)=
\alpha'^{n-3}g^{n-2}
\int_{0}^{1}\frac{dz_{n-1}}{z_{n-1}}\dots\int_{0}^{1}\frac{dz_{3}}{z_{3}}\int_{0}^{1}\frac{dz_{2}}{z_{2}}\Bigl\langle\hat{f}\Bigr|\Bigl(z_{n-1}^{L_{0}-1}V_{n-1}(1)z_{n-1}^{-(L_{0}-1)}\Bigr)\cr
\Bigl((z_{n-1}z_{n-2})^{L_{0}-1}V_{n-1}(1)(z_{n-1}z_{n-2})^{-(L_{0}-1)}\Bigr)\dots
 \Bigl((z_{3}z_{4}\dots z_{n-1})^{L_{0}-1}V_{3}(1)(z_{3}z_{4}\dots
 z_{n-1})^{-(L_{0}-1)}\Bigr)\cr
\times\Bigl((z_{2}z_{3}\dots z_{n-1})^{L_{0}-1}V_{2}(1)(z_{2}z_{3}\dots z_{n-1})^{-(L_{0}-1)}\Bigr)\Bigl|\hat{i}\Bigr\rangle\,.
\end{multline}
A similar change of variables
\begin{align}
y_{2} & := z_{2}z_{3}z_{4}\dots z_{n-1}\,\nonumber \\
y_{3} & := z_{3}z_{4}\dots z_{n-1}\,\nonumber \\
 & \vdots\nonumber \\
y_{n-2} & := z_{n-2}z_{n-1} \,\nonumber \\
y_{n-1} & := z_{n-1}\,,
\end{align}
this time leads to $(n-2)$ instead of $(n-3)$ vertex operators to
be integrated between $(0,1)$,
\begin{equation}
\mathcal{J}_{n}(123\dots n)=(\alpha'g)^{n-2}\int_{0<y_{2}<y_{3}<\dots<y_{n-1}<1}\prod_{i=2}^{n-2}dy_{i}\,\Bigl\langle\hat{f}\Bigr|\frac{V_{n-1}(y_{n-1})}{y_{n-1}}\,\frac{V_{n-2}(y_{n-2})}{y_{n-2}}\dots\frac{V_{3}(y_{3})}{y_{3}}\,\frac{V_{2}(y_{2})}{y_{2}}\Bigl|\hat{i}\Bigr\rangle.\label{eq:current}
\end{equation}
As discussed earlier in this paper, in light of the
fact that field theory momentum kernel happens to be the inverse of
propagator matrix in an$(n-2)!$ basis, 
the (field theory) numerator can be calculated from the $\alpha'$
limit of the following sum over $(n-2)!$ permutations of the legs
$\{2,3,4,\dots,n-1\}$,
\begin{equation}
n(123\dots n)=\lim_{k_{n}^{2},\alpha'\rightarrow0}\,\sum_{\beta\in S_{n-2}}\mathcal{S}_{\alpha'}[n-1,\dots,4,3,2|\beta]\,\mathcal{J}_{n}(1,\beta,n).\label{eq:n-2-string-numerator}
\end{equation}
Despite perhaps a bit complicated at first sight, it is known \cite{Du:2011js,Du:2016tbc}
that the above permutation sum very often simplifies if we organise the
amplitudes (or currents) in the equation according to the so-called Fundamental  BCJ 
relations~\cite{BjerrumBohr:2009rd,Stieberger:2009hq,Feng:2010my}, 

\begin{align}
n(123\dots n)= & \lim_{k_{n}^{2},\alpha'\rightarrow0}\,\sum_{\beta\in S_{n-3}}\mathcal{S}_{\alpha'}[n-2,\dots,4,3,2|\beta]\,\Bigl\{ 2i\sin(\pi\alpha'k_{n-1}\cdot k_{1})\mathcal{J}_{n}(1,n-1,\beta_{2},\beta_{3},\dots,\beta_{n-2},n)\nonumber \\
 & +2i\sin(\pi\alpha'k_{n-1}\cdot(k_{1}+k_{\beta_{2}}))\mathcal{J}_{n}(1,\beta_{2},n-1,\beta_{3},\dots,\beta_{n-2},n)\nonumber \\
 & +\dots\nonumber \\
 & +2i\sin(\pi\alpha'k_{n-1}\cdot(k_{1}+k_{\beta_{2}}+\dots+k_{\beta_{n-2}}))\mathcal{J}_{n}(1,\beta_{2},\beta_{3},\dots,\beta_{n-2},n-1,n)\Bigr\}.
\end{align}
Explicitly we focus first on the subset of the full $(n-2)!$ permutation
sum in equation (\ref{eq:n-2-string-numerator}) where leg $(n-1)$
is inserted between the relatively fixed set $\{2,3,4,\dots,n-2\}$.
It is straightforward to see from the definition of momentum kernel
that this is proportional to
\begin{align}
 & 2i\sin(\pi\alpha'k_{n-1}\cdot k_{1})\mathcal{J}_{n}(1,n-1,2,3,\dots,n-2,n)+2i\sin(\pi\alpha'k_{n-1}\cdot(k_{1}+k_{2}))\mathcal{J}_{n}(1,2,n-1,3,\dots,n-2,n)\nonumber \\
 &+\dots +2i\sin(\pi\alpha'k_{n-1}\cdot(k_{1}+k_{2}+\dots+k_{n-2}))\mathcal{J}_{n}(1,2,3,\dots,n-2,n-1,n)\,.\label{eq:bcj-sum}
\end{align}
To complete this calculation we then perform an $(n-3)!$ permutation
sum on the originally fixed set $\{2,3,4,\dots,n-2\}$, which in turn
can be organised into BCJ sums if we regard it as insertions of the
leg $(n-2)$ between set $\{2,3,4,\dots,n-3\}$ and so on. We will
see in the next section that algebraic structure can be identified
in this process. We shall also drop the distinction between $\Bigl|\hat{f}\Bigr\rangle$
and $\Bigl|f\Bigr\rangle$ since it is known that at least in the
field theory limit, numerators obtained through equation (\ref{eq:n-2-qft-numerator})
remain finite in the on-shell limit. 

\section{Half-ladder basis numerators and the vertex operator algebra}
\label{sec:the-algebra}

The BCJ sum (\ref{eq:bcj-sum}) written down in the last section can
be more easily understood in the language of vertex operators. We
will in the following illustrate this idea through two lower-point
examples. For simplicity we will assume the scattering particles are
all tachyons, although the generalisation to gluons is straightforward. 

\subsection{The three point numerator}
\label{sec:the-algebra-3pt}

\begin{figure}[t]
\centering
\includegraphics[width=8cm]{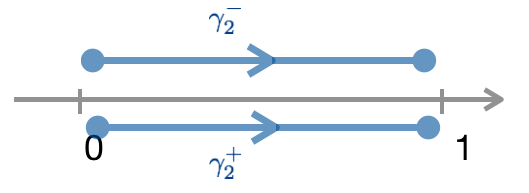}
\label{Fig:gamma-2}
\caption{Integration contours present in the evaluation of three point kinematic
numerator}
\end{figure}

\begin{figure}[t]
\centering
\subfigure[]{
\includegraphics[width=8cm]{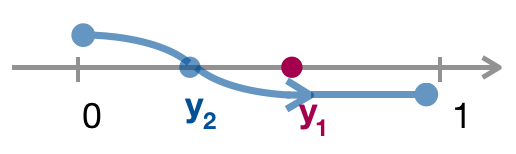}
 }
\subfigure[]{
\includegraphics[width=8cm]{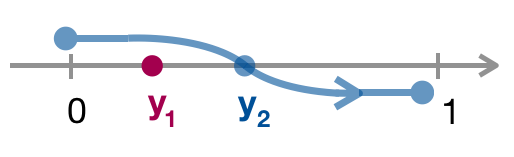}
}
\caption{The contours corresponding to $V_{1}V_{2}$ and $V_{2}V_{1}$ respectively. }
\label{Fig:analytic-continuations}
\end{figure}

Consider the numerator at three points. Equation (\ref{eq:n-2-string-numerator})
reads
\begin{equation}
2i\,\sin(\pi\alpha'k_{2}\cdot k_{1})g\alpha'\int_{0}^{1}\frac{dy_{2}}{y_{2}}\,y_2^{\alpha'k_{2}\cdot k_{1}}=g\alpha'\int_{\gamma_{2}^{+}-\gamma_{2}^{-}}\frac{dy_{2}}{y_{2}}\,\left(-y_{2}\right)^{\alpha'k_{2}\cdot k_{1}}\,,\label{eq:3pt-eqn1}
\end{equation}
where we followed the same reasoning used in the extraction of string
momentum kernel from monodromy relations \cite{BjerrumBohr:2010hn} to translate
the sine factor into the difference between integrals $\gamma_{2}^{+}$
and $\gamma_{2}^{-}$ along opposite sides of a branch cut (Fig \ref{Fig:gamma-2}).
To see how equation (\ref{eq:3pt-eqn1}) can  have an algebraic
origin, recall that in the operator language a Veneziano type world-sheet
integral formula derives from normal ordering two vertex operators
\begin{align}
V(y_{1})V(y_{2}) &
                   =\left(e^{\sqrt{\alpha'}\sum_{1}^{\infty}\frac{k_{1}\cdot\alpha_{-n}}{n}y_{1}^{n}}e^{-\sqrt{\alpha'}\sum_{1}^{\infty}\frac{k_{1}\cdot\alpha_{n}}{n}y_{1}^{-n}}e^{ik_{1}\cdot
                   x}e^{k_{1}\cdot p\,\ln
                   y_{1}}\right)\cr
                   &\times\left(e^{\sqrt{\alpha'}\sum_{1}^{\infty}\frac{k_{2}\cdot\alpha_{-n}}{n}y_{2}^{n}}e^{-\sqrt{\alpha'}\sum_{1}^{\infty}\frac{k_{2}\cdot\alpha_{n}}{n}y_{2}^{-n}}e^{ik_{2}\cdot
                   x}e^{k_{2}\cdot p\,\ln y_{2}}\right)\nonumber \\
 & =e^{\alpha'k_{1}\cdot k_{2}\,\ln y_{1}}e^{-\alpha'\sum_{1}^{\infty}\frac{1}{n}\left(\frac{y_{2}}{y_{1}}\right)^{n}}:V(y_{1})V(y_{2}):\nonumber \\
 & \sim e^{\alpha'k_{1}\cdot k_{2}\,\ln(y_{1}-y_{2})}f(y).
\end{align}
Starting from a point $y_{2}$ on the real line with $\Bigl|\frac{y_{2}}{y_{1}}\Bigr|<1$
so that the infinite series converges $-\sum_{1}^{\infty}\frac{1}{n}\left(\frac{y_{2}}{y_{1}}\right)^{n}=\ln(1-\frac{y_{2}}{y_{1}})$,
the analytic continuation of the product $V(y_{1})V(y_{2})\sim e^{\alpha'k_{1}\cdot k_{2}\,\ln(y_{1}-y_{2})}f(y)$
is uniquely determined as $y_{2}$ travels continuously on the complex
plane. Bearing this in mind, it is straightforward to see that integration
contour associated with the ordered product 
\begin{equation}
\frac{V(y_{1})}{y_{1}}\,\left(\int_{0}^{1}dy_{2}\frac{V(y_{2})}{y_{2}}\right)\,,
\end{equation}
analytically continues as illustrated in Fig \ref{Fig:analytic-continuations}(a),
while the contour associated with opposite order
\begin{equation}
\left(\int_{0}^{1}dy_{2}\frac{V(y_{2})}{y_{2}}\right)\,\frac{V(y_{1})}{y_{1}}\,,
\end{equation}
corresponds to Fig \ref{Fig:analytic-continuations}(b). (In the later
case $V(y_{2})V(y_{1})\sim e^{\alpha'k_{1}\cdot k_{2}\,\ln(y_{2}-y_{1})}f(y)$
and the branch cut lies along the $y_{2}<y_{1}$ part of the real
line.)\footnote{Here we are assuming all the $y_{i}$'s were  shifted initially in
the imaginary direction $y_{i}\rightarrow y_{i}+i\epsilon\delta_{i}$
as in \cite{BjerrumBohr:2010hn} prior to the analytic continuation along
the horizontal line $(0+i\epsilon\delta_{i},1+i\epsilon\delta_{i})$,
along which the integration is performed. The shifts $\delta_{i}$
were defined according to the ordering of the vertex operators so
that $\delta_{i}<\delta_{j}$ if operator $V_{j}$ appears on the
right side of $V_{i}$. The integration contours become those shown
in Fig \ref{Fig:analytic-continuations} when $\delta\rightarrow0$.}

A careful inspection on the three-point numerator obtained earlier
in (\ref{eq:3pt-eqn1}) then shows that the numerator can be written
in the operator language as  
\begin{align}
 & \alpha'\int_{\gamma_{2}^{+}}\frac{dy_{2}}{y_{2}}\,\left(-y_{2}\right)^{\alpha'k_{2}\cdot k_{1}}-e^{-i\pi\alpha'k_{2}\cdot k_{1}}\alpha'\int_{\gamma_{2}^{-}}\frac{dy_{2}}{y_{2}}\,y_{2}^{\alpha'k_{2}\cdot k_{1}}\nonumber \\
 & =\alpha'\left(\int_{0}^{1}dy_{2}\Bigl\langle f\Bigr|\frac{V_{1}(y_{1})}{y_{1}}\,\frac{V_{2}(y_{2})}{y_{2}}\Bigl|0\Bigr\rangle-e^{-i\pi\alpha'k_{2}\cdot k_{1}}\int_{0}^{1}dy_{2}\Bigl\langle f\Bigr|\frac{V_{2}(y_{2})}{y_{2}}\,\frac{V_{1}(y_{1})}{y_{1}}\Bigl|0\Bigr\rangle\right)\Bigr|_{y_{1}\rightarrow0}\nonumber \\
 & =\alpha'\Bigl\langle f\Bigr|[T_{1},T_{2}]_{\alpha'}\Bigl|0\Bigr\rangle\,,
\end{align}
 where $\langle f|$ is the external state
  defined~\eqref{e:modified} for $n=3$.
This is in some sense the structure we anticipated. However instead
of a Lie bracket, we arrive at a similar, yet asymmetric structure
defined as\footnote{The bracket observed here can be identified as a weakly closed system.
Given an associative algebra $\mathbb{A}$ over some field $F$, a
weakly closed system is the closed set of elements under an additional
multiplicative operation (bracket) $\times:\mathbb{A}\otimes\mathbb{A}\rightarrow\mathbb{A}$
of the form
\begin{equation}
a\times b=ab+\gamma(a,b)ba
\end{equation}
For some $\gamma(a,b)\in F$. In principle it is actually possible to restore Lie brackets by introducing co-cycles
similar to those discussed in~\cite[section~6.4]{Green:1987sp}. In this picture 
additional order dependent exponential factors factorise from the numerator.
}
\begin{equation}
[T_{1},T_{2}]_{\alpha'}=T_{1}T_{2}-e^{-i\pi\alpha'k_{2}\cdot k_{1}}T_{2}T_{1}\label{eq:asymm-bracket}
\end{equation}
where we introduced the shorthand notation to denote $T_{2}=\int_{0}^{1}dy_{2}\frac{V_{2}(y_{2})}{y_{2}}$
and $T_{1}=\frac{V_{1}(y_{1})}{y_{1}}\Bigr|_{y_{1}\rightarrow0}$.

\subsection{Four points and higher}

\label{sec:the-algebra-4pt}

The numerator at four points follows similar calculations. As discussed
earlier we note that BCJ sums can be extracted from the full $(n-2)!$
permutation sum appears in (\ref{eq:n-2-string-numerator}).
\begin{align}
 & 2i\,\sin(\pi\alpha'k_{3}\cdot k_{1})\mathcal{J}_{4}(1324)+2i\,\sin(\pi\alpha'k_{3}\cdot(k_{1}+k_{2}))\mathcal{J}_{4}(1234)\nonumber \\
 & =2i\,\sin(\pi\alpha'k_{3}\cdot k_{1})g^2{\alpha'}^2\int_{0}^{y_{2}}\frac{dy_{3}}{y_{3}}\,y_{3}^{\alpha'k_{3}\cdot k_{1}}(y_{2}-y_{3})^{\alpha'k_{3}\cdot k_{2}}f(y_{3})\nonumber \\
 & +2i\,\sin(\pi\alpha'k_{3}\cdot(k_{1}+k_{2})) g^2{\alpha'}^2\int_{y_{2}}^{1}\frac{dy_{3}}{y_{3}}\,y_{3}^{\alpha'k_{3}\cdot k_{1}}(y_{3}-y_{2})^{\alpha'k_{3}\cdot k_{2}}f(y_{3})\nonumber \\
 & =g^2{\alpha'}^2\int_{\gamma_{3}^{+}-\gamma_{3}^{-}}\frac{dy_{3}}{y_{3}}\,\left(-y_{3}\right)^{\alpha'k_{3}\cdot k_{1}}(y_{2}-y_{3})^{\alpha'k_{3}\cdot k_{2}}f(y_{3})\,,\label{eq:4pt-eqn1}
\end{align}
where in the last line of the equation above we translated the phase
factors introduced by sines into the difference between integrals
$\gamma_{3}^{+}$ and $\gamma_{3}^{-}$ below and above the branch
cut, similar to those illustrated in Fig \ref{Fig:gamma-2}. The first
and the second term contribute the $0<\Bigl|y_{3}\Bigr|<y_{2}$ and
$y_{2}<\Bigl|y_{3}\Bigr|<1$ segment of the $\gamma_{3}^{+}$ contour
respectively, and likewise for the contour $\gamma_{3}^{-}$. The
integration in equation
(\ref{eq:4pt-eqn1}) can be subsequently rewritten as
\begin{align}
 & \int_{\gamma_{3}^{+}}\frac{dy_{3}}{y_{3}}\,\left(-y_{3}\right)^{\alpha'k_{3}\cdot k_{1}}(y_{2}-y_{3})^{\alpha'k_{3}\cdot k_{2}}f(y_{3})-e^{i\pi\alpha'k_{3}\cdot(k_{1}+k_{2})}\int_{\gamma_{3}^{-}}\frac{dy_{3}}{y_{3}}\,y_{3}^{\alpha'k_{3}\cdot k_{1}}(y_{3}-y_{2})^{\alpha'k_{3}\cdot k_{2}}f(y_{3})\nonumber \\
 & =\Bigl\langle f\Bigr|\frac{V(y_{2})}{y_{2}}\,\frac{V(y_{1})}{y_{1}}\,\left(\int_{0}^{1}dy_{3}\frac{V(y_{3})}{y_{3}}\right)\Bigl|0\Bigr\rangle-e^{i\pi\alpha'k_{3}\cdot(k_{1}+k_{2})}\Bigl\langle f\Bigr|\left(\int_{0}^{1}dy_{3}\frac{V(y_{3})}{y_{3}}\right)\,\frac{V(y_{2})}{y_{2}}\,\frac{V(y_{1})}{y_{1}}\Bigl|0\Bigr\rangle\,,
\end{align}
again here  $\langle f|$ is the external state
  defined~\eqref{e:modified} for $n=4$.
What remains in the calculation is a multiplication by $2i\,\sin(\pi\alpha'k_{2}\cdot k_{1})$,
which has the same effect as was shown at three points, and we arrive
at
\begin{equation}
n(1234)=g^2{\alpha'}^2\lim_{k_{4}^{2},\alpha'\rightarrow0}\Bigl\langle f\Bigr|[[T_{1},T_{2}]_{\alpha'},T_{3}]_{\alpha'}\Bigl|0\Bigr\rangle\,,
\end{equation}
where $T_{2}=\int_{0}^{1}dy_{2}\frac{V_{2}(y_{2})}{y_{2}}$, $T_{3}=\int_{0}^{1}dy_{3}\frac{V_{3}(y_{3})}{y_{3}}$
and $T_{1}=\frac{V_{1}(y_{1})}{y_{1}}\Bigr|_{y_{1}\rightarrow0}$.
Generically this translation procedure continues, and the $n$-point
half-ladder basis numerator is given by

\begin{equation}
n(123\dots n)=g^{n-2}{\alpha'}^{n-2}\lim_{k_{n}^{2},\alpha'\rightarrow0}\Bigl\langle f\Bigr|[[[T_{1},T_{2}]_{\alpha'},T_{3}]_{\alpha'}\dots,T_{n-1}]_{\alpha'}\Bigl|0\Bigr\rangle.\label{eq:npt-numerator}
\end{equation}

Incidentally the same $\alpha'$-weighted commutator structure was
actually observed earlier in \cite{Ma:2011um} from a rather
different perspective, and more recently in \cite{Carrasco:2016ygv}.
There it was shown that, starting from the standard colour-order formulation
and applying the Kleiss-Kuijf relations~\cite{Kleiss:1988ne} on the colour-ordered open string amplitudes,
the result is a formula bearing much resemblance to the Del Duca-Dixon-Maltoni
expression, except with the Lie bracket between the $SU(N)$ colour
generators $t_{a}$ replaced by exactly the same $\alpha'$-weighted
commutator (\ref{eq:asymm-bracket}).
\begin{align}
\mathcal{M}_{n} & =\sum_{\rho\in S_{n-1}}tr(t_{\rho(1)}t_{\rho(2)}\dots t_{\rho(n-1)}t_{n})\,\mathcal{A}_{n}(\rho(1),\rho(2),\dots,\rho(n-1),n)\nonumber \\
 & =\sum_{\sigma\in S_{n-2}}tr([[[t_{1},t_{\sigma(2)}]_{\alpha'},t_{\sigma(3)}]_{\alpha'}\dots,t_{\sigma(n-1)}]_{\alpha'}t_{n})\,\mathcal{A}_{n}(1,\sigma(2),\dots,\sigma(n-1),n)\,.\label{eq:open-string-ddm}
\end{align}
In this sense we see that the numerator structure (\ref{eq:npt-numerator})
just derived serves as the colour-kinematics counterpart of (\ref{eq:open-string-ddm})
in the context of string theory.

Note that the asymmetry feature of the new bracket $[\,,\,]_{\alpha'}$
is entirely introduced by the phase factor $e^{-i\pi\alpha'k_{1}\cdot k_{2}}$
which vanishes in the $\alpha'\rightarrow0$ limit, so that as far
as field theory is concerned, we might as well replace all bracket with
Lie bracket in the (field theory) numerator formula. 
In the section~\ref{sec:explicit-generators} we will work out some explicit examples.
The basis numerator $n(123\dots n)$ thus obtained apparently respects Jacobi identity.
Under this setting, a generic (not necessarily half-ladder) numerator
is then simply given by the commutator structure suggested by its
associated cubic graph, for example 
\begin{equation}
\begin{minipage}{4.2cm}
\includegraphics[width=4.2cm]{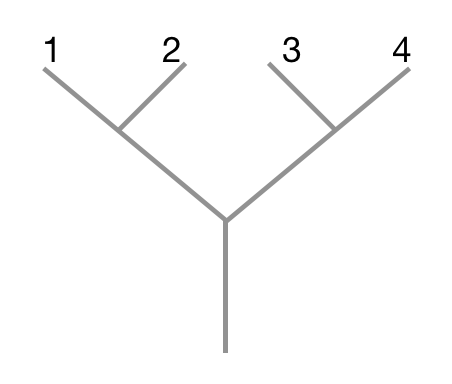}
\end{minipage}
\sim[[T_{1},T_{2}],[T_{3},T_{4}]]
\end{equation}

\subsection{Explicit generators }

\label{sec:explicit-generators}

Despite the apparent formal simplicity attained by expressing kinematic
numerator in the language of vertex operator algebra, it is not yet
clear to us whether the generator $T_{i}=\int_{0}^{1}dy\frac{V_i(y)}{y}$
has a simple representation so that the numerators can be calculated
from the algebra directly. Instead, in this section we carry out the
integral term-wise. Hopefully the calculation can provide useful
insight regarding the analytic behaviour of the generators. 

As an illustration of how this proceeds, for the moment we make a
bit of digression and calculate the three-point numerator directly
instead of calculating $T_{i}$, which takes the same generic form
as far as the integral is concerned, except simpler. The four-point
case requires the evaluation of the disc  integral and will be treated in section~\ref{sec:hyper-geometric-functions}. Taking into account
that momentum kernel at three points is simply a sine factor, and that
the current is given by equation (\ref{eq:current}), the numerator
defined in (\ref{eq:n-2-string-numerator}) reads
\begin{equation}
n(123)=\lim_{k_{3}^{2},\alpha'\rightarrow0}\sin(\pi\alpha'k_{2}\cdot k_{1})g\alpha'\Bigl\langle\tilde{f}\Bigr|:\int_{0}^{1}\frac{dy_{2}}{y_{2}}\epsilon_{2}\cdot X(y_{2})e^{ik_{2}\cdot X}:\,\Bigl|\tilde{i}\Bigr\rangle.
\end{equation}
Plugging the explicit mode expansion of the vertex operator shows
that the expectation value contributes the following integral.
\begin{equation}
\Bigl\langle\tilde{f}\Bigr|:\int_{0}^{1}\frac{dy_{2}}{y_{2}}\epsilon_{2}\cdot X(y_{2})e^{ik_{2}\cdot X}:\,\Bigl|\tilde{i}\Bigr\rangle= V_{3}\,\int_{0}^{1}dy_{2}\frac{1}{y_{2}}e^{\alpha'k_{2}\cdot
   k_{1}\,\ln y_{2}}\,,
\end{equation}
where $V_{3}$ is the Yang-Mills cubic
\begin{equation}
V_{3}=\left(\epsilon_{2}\cdot\epsilon_{1}\right)\left(k_{2}\cdot\epsilon_{3}\right)+\left(\epsilon_{3}\cdot\epsilon_{2}\right)\left(k_{2}\cdot\epsilon_{1}\right)+\left(\epsilon_{3}\cdot\epsilon_{1}\right)\left(k_{2}\cdot\epsilon_{1}\right)\,.
\end{equation}
Namely in this case the world-sheet integral factorises, and gives
\begin{equation}
\int_{0}^{1}dy_{2}\,\left(y_{2}\right)^{\alpha'k_{2}\cdot k_{1}-1}=\frac{1}{\alpha'k_{2}\cdot k_{1}}\,.
\end{equation}
The three-point current is therefore $V_{3}/k_{3}^{2}$ as expected.
Multiplying current by momentum kernel $\sin(\pi\alpha'k_{2}\cdot k_{1})\sim\pi\alpha'k_{2}\cdot k_{1}$
shows that the numerator is $V_{3}$ in the $\alpha'$ limit, which
is consistent with known result.

The integral involved in calculating the $T_{a}$'s is similar. Assuming
vertex operator has the following expansion,

\begin{equation}
V(y)=\sum_{n=-\infty}^{\infty}e^{ik\cdot x}a_{n}y^{n}e^{\sqrt{\alpha'} k\cdot p\,\ln\,y}\,,
\end{equation}
where $a_{n}$ are expansion coefficients, the integration of interest
$\int_{0}^{1}dy\frac{V(y)}{y}$ can be readily carried out term-wise.
\begin{align}
\int_{0}^{1}\frac{dy}{y}V(y) & =\sum_{n=-\infty}^{\infty}e^{ik\cdot x}a_{n}\frac{1}{\sqrt{\alpha'} k\cdot\alpha_{0}+n}\,.\label{eq:operator}
\end{align}
The expansion coefficients $a_{n}$ are not so difficult to compute
either. Keeping only the first few power terms in $\sqrt{\alpha'}$,
we have
\begin{align}
a_{0}= & \dots+\epsilon\cdot\alpha_{1}\left(\sqrt{\alpha'} k\cdot\alpha_{-1}-\sqrt{\alpha'}^{2}\sum_{n=1}^{\infty}\left(\frac{k\cdot\alpha_{-(n+1)}}{n+1}\right)\left(\frac{k\cdot\alpha_{n}}{n}\right)+\mathcal{O}(\sqrt{\alpha'}^{3})\right)\nonumber \\
 & +\epsilon\cdot\alpha_{0}\left(1-\sqrt{\alpha'}^{2}\sum_{n=1}^{\infty}\left(\frac{k\cdot\alpha_{-n}}{n}\right)\left(\frac{k\cdot\alpha_{n}}{n}\right)+\mathcal{O}(\sqrt{\alpha'}^{3})\right)\nonumber \\
 & +\epsilon\cdot\alpha_{-1}\left(-\sqrt{\alpha'} k\cdot\alpha_{1}-\sqrt{\alpha'}^{2}\sum_{n=1}^{\infty}\left(\frac{k\cdot\alpha_{-n}}{n}\right)\left(\frac{k\cdot\alpha_{n+1}}{n+1}\right)+\mathcal{O}(\sqrt{\alpha'}^{3})\right)+\dots\,,\label{eq:a0}
\end{align}
and similarly for the rest of the $a_{n}$.

\section{Structure constants of the vertex operator algebra}

In light of the vertex operator expression of the BCJ numerators, a
natural question to ask is whether the algebraic structure observed
can be identified with any existing algebra. 
Note especially in 
\cite{Green:1987sp}
this was done in a similar setting to explain the $E_{8}\times E_{8}$ symmetry
of the heterotic string theory, in which case all  momentum inner product 
$k_{i}\cdot k_{j}$ appear in the exponent take integer values due to the
compactification so that the world-sheet integrals can be 
elegantly performed.
In the presence of branch cuts the story gets considerably messier.
We saw in the last section
that the generators can be calculated in a term-by-term basis, in
this sense it is part of the universal enveloping algebra $U(g)$
of Virasoro modes, which is however too general to provide any useful
information. On the other hand it is known that the self-dual sector
of Yang-Mills numerator is explained by area-preserving diffeomorphism
algebra  as exploited in~\cite{Monteiro:2011pc}. We shall see in the following discussion that in the field theory limit, the vector
$\times$ vector $\rightarrow$ vector sector is indeed isomorphic
to diffeomorphism. We will see that the $\alpha'$-weighted commutator
(\ref{eq:asymm-bracket}) produces the familiar structure constant
\begin{equation}
f_{ab}{}^{c}=-i\left(\delta_{a}{}^{c}k_{1b}-\delta_{b}{}^{c}k_{2a}\right)\,,\label{eq:structure-const}
\end{equation}
up to $\alpha'$ corrections.

\subsection{Reproducing the diffeomorphism algebra}

\label{sec:the-diffeo}
\begin{figure}[t]
\centering
\subfigure[]{
\includegraphics[width=7cm]{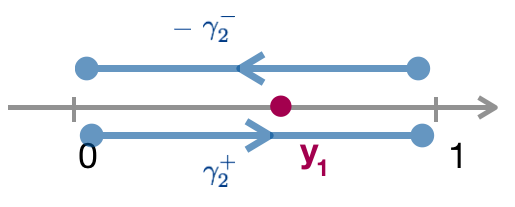}
\label{fig:diffeo1}
}
\subfigure[]{
\includegraphics[width=7cm]{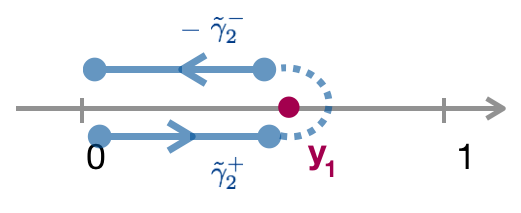}
\label{fig:diffeo2}
}
\caption{The contours present in the $\alpha'$-weighted commutator calculation}
\end{figure}
For simplicity we write the vertex operator for vector particle as
$V(y)=e^{ik\cdot X+\epsilon\cdot\dot{X}}$, bearing in mind that all
calculations in the following are kept only to the first order in
$\epsilon_{i}$'s. It is then straightforward to compute the product
of two vertex operators using standard Wick contractions. Recall that
the integration contours associated with the products $V_{1}(y_{1})V_{2}(y_{2})$
and $V_{2}(y_{2})V_{1}(y_{1})$ correspond to Fig \ref{Fig:analytic-continuations}(a)
and (b) respectively. Taking a relative minus sign and phase into
account, we see that the $\alpha'$-weighted commutator is given by
the following formula, integrated over the contour $\gamma^{+}-\gamma^{-}$shown
in Fig \ref{fig:diffeo1}, or equivalently, over the arc surrounding
$y_{1}$ plus the difference $\tilde{\gamma}_{2}^{+}-\tilde{\gamma}_{2}^{-}$
along two sides of the branch cut $(0,y_{1})$ (Fig~\ref{fig:diffeo2}).
\begin{align}
\int_{0}^{1}\frac{dy_{2}}{y_{2}}[V_{1}(y_{1}),V_{2}(y_{2})]_{\alpha'}= & \int_{\text{around }y_{1}}+\int_{\tilde{\gamma}^{+}-\tilde{\gamma}^{-}}\frac{dy_{2}}{y_{2}}\,e^{\alpha'k_{1}\cdot k_{2}\ln(y_{1}-y_{2})}\nonumber \\
 & \left(1-\alpha'\epsilon_{2}\cdot k_{1}\frac{y_{2}}{y_{1}-y_{2}}+\alpha'\epsilon_{1}\cdot k_{2}\frac{y_{1}}{y_{1}-y_{2}}+\alpha'\epsilon_{1}\cdot\epsilon_{2}\frac{y_{1}y_{2}}{(y_{1}-y_{2})^{2}} \right. 
 \nonumber \\
&
 \left. -\alpha'^{2}(\epsilon_{1}\cdot k_{2})(\epsilon_{2}\cdot k_{1})\frac{y_{1}y_{2}}{(y_{1}-y_{2})^{2}}\right)
  :V_{1}(y_{1})V_{2}(y_{2}):\Bigr|_{\text{linear in }\epsilon}\,.\label{eq:bracket-v2v1}
\end{align}
The $\epsilon_{1}\cdot\epsilon_{2}$ term can be made to vanish with
suitable gauge choice or for a self-dual
Yang-Mills amplitude, therefore we know that it is not relevant to
the structure of interest here; while the $(\epsilon_{1}\cdot k_{2})(\epsilon_{2}\cdot k_{1})$
term carries an $\alpha'^{2}$ and becomes sub-leading in the field
theory limit. We have checked that the construction generalises to the
superstring in similar fashion as the supersymmetric Berends-Giele  
current of~\cite[App.~A]{Berg:2016fui}.
For the moment let us focus first on the $\epsilon_{1}\cdot k_{2}$
and $\epsilon_{2}\cdot k_{1}$ terms. We will soon see that these
two terms reproduce the familiar diffeomorphism structure constant
multiplied by a vector (spin~1) vertex operator, whereas the factor $1$ contributes  the tensor part  and the 
$\epsilon_{1}\cdot\epsilon_{2}$ a scalar (spin 0) needed to reproduce the quartic vertex
contribution in the field theory limit.

For generic values of $k_{1}\cdot k_{2}$ the integral around $y_{1}$
vanishes as the radius of the arc $\varepsilon$ approaches zero.
This is because the normal ordered product $:V_{1}(y_{1})V_{2}(y_{2}):$
contains no pole at $y_{1}$ so that when expanded as power series
of $(y_{2}-y_{1})$, the integral becomes\footnote{Note that the branch cut of $\ln\,y_{2}$ lies only along $y_{2}<0$,
so that it can be safely expanded as well into a power series in $(y_{2}-y_{1})$
without producing any pole at $y_{1}$. $\ln\,y_{2}=\ln\left(y_{1}+(y_{2}-y_{1})\right)=\ln\,y_{1}+\frac{1}{y_{1}}(y_{2}-y_{1})+\dots$.}

\begin{equation}
\int_{-\pi}^{\pi}\varepsilon e^{i\theta}id\theta\frac{1}{\varepsilon e^{i\theta}}\,e^{\alpha'k_{1}\cdot k_{2}(\ln\varepsilon+i\theta)}\Bigl(c_{0}+c_{1}\varepsilon e^{i\theta}+\dots\Bigr)\xrightarrow{\varepsilon\rightarrow0}\,0\,.
\end{equation}

Similarly, we consider formally expanding $:V_{1}(y_{1})V_{2}(y_{2}):$
in the $\tilde{\gamma}^{+}$ and $\tilde{\gamma}^{-}$ part of the
integral, and note that an $n$-th power term contributes as
\begin{align*}
\int_{\tilde{\gamma}_{2}^{\pm}}dy_{2}(y_{2}-y_{1})^{n}\,e^{\alpha'k_{1}\cdot k_{2}\ln(y_{1}-y_{2})} & =e^{i\pi\alpha'k_{1}\cdot k_{2}}\int_{0\mp i\varepsilon}^{y_{1}\mp i\varepsilon}dy_{2}(y_{2}-y_{1})^{n+\alpha'k_{1}\cdot k_{2}}\\
 & =\frac{1}{n+\alpha'k_{1}\cdot k_{2}+1}e^{i\pi\alpha'k_{1}\cdot k_{2}}\left[0-e^{\mp i\pi\alpha'k_{1}\cdot k_{2}}e^{\alpha'k_{1}\cdot k_{2}\ln\,y_{1}}\right].
\end{align*}
For generic integer $n$ the difference $\tilde{\gamma}_{2}^{+}-\tilde{\gamma}_{2}^{-}$
then yields $\frac{2i\,\sin(\pi\alpha'k_{1}\cdot k_{2})}{n+\alpha'k_{1}\cdot k_{2}+1}e^{\alpha'k_{1}\cdot k_{2}\ln\,y_{1}}$,
which also vanishes as $\alpha'\rightarrow0$. The only exception
occurs when $n=-1$, in which case the difference $\tilde{\gamma}_{2}^{+}-\tilde{\gamma}_{2}^{-}$
returns a finite result,
\begin{equation}
\frac{2i\,\sin(\pi\alpha'k_{1}\cdot k_{2})}{\alpha'k_{1}\cdot k_{2}}e^{\alpha'k_{1}\cdot k_{2}\ln\,y_{1}}=2\pi i+\mathcal{O}(\alpha').
\end{equation}
In other words the effect of the branch cut integration and the $\alpha'$
limit combine is equivalent to stripping off an exponential phase
factor $e^{\alpha'k_{1}\cdot k_{2}\ln(y_{1}-y_{2})}$ and picking up
the $-1$ power term coefficient if we imagine the rest of the integrand
$\left(-\alpha'\epsilon_{2}\cdot k_{1}\frac{y_{2}}{y_{1}-y_{2}}+\alpha'\epsilon_{1}\cdot k_{2}\frac{y_{1}}{y_{1}-y_{2}}\right):V_{1}(y_{1})V_{2}(y_{2}):$
being expressed as a holomorphic series in $(y_{2}-y_{1})$. The procedure
just described can be recast in the language of a Cauchy integral
if we would like, writing the commutator as
\begin{align}
\int_{0}^{1}\frac{dy_{2}}{y_{2}}[V_{1}(y_{1}),V_{2}(y_{2})]_{\alpha'}=
& \oint_{y_{1}}dy_{2}\alpha'(-\alpha'\epsilon_{2}\cdot k_{1}\frac{y_{2}}{y_{1}-y_{2}}+\alpha'\epsilon_{1}\cdot k_{2}\frac{y_{1}}{y_{1}-y_{2}}):V_{1}(y_{1})V_{2}(y_{2}):\Bigr|_{\text{linear in }\epsilon} \nonumber \\
& +\text{scalars and tensors}\,,
\end{align}
up to $\alpha'$ corrections, and the result is the same as replacing
all $y_{2}$ dependence in the numerator by $y_{1}$ (taking reference
to Cauchy residue theorem). Plugging in the explicit form of vector
(spin~1)
vertex operator this gives
\begin{align}
 & (-\alpha'\epsilon_{2}\cdot k_{1}\frac{y_{2}}{y_{1}-y_{2}}+\alpha'\epsilon_{1}\cdot k_{2}\frac{y_{1}}{y_{1}-y_{2}}):\epsilon_{1}\cdot\dot{X}(y_{1})e^{ik_{1}\cdot X(y_{1})}\;\epsilon_{2}\cdot\dot{X}(y_{2})e^{ik_{2}\cdot X(y_{2})}:\Bigr|_{\begin{array}{c}
y_{2}\rightarrow y_{1}\\
\text{linear in }\epsilon
\end{array}}\nonumber \\
 & =-\left(\epsilon_{2}\cdot k_{1}\right)\epsilon_{1}\cdot\dot{X}e^{i(k_{1}+k_{2})\cdot X(y_{1})}+\left(\epsilon_{1}\cdot k_{2}\right)\epsilon_{2}\cdot\dot{X}e^{i(k_{1}+k_{2})\cdot X(y_{1})},
\end{align}
and we see that the piece proportional to a vector (spin~1) vertex operator is
indeed isomorphic to the diffeomorphism algebra as expected
\begin{equation}
\int_{0}^{1}\frac{dy_{2}}{y_{2}}[\epsilon_{1}\cdot\dot{X}(y_{1})e^{ik_{1}\cdot X(y_{1})},\,\epsilon_{2}\cdot\dot{X}(y_{2})e^{ik_{2}\cdot X(y_{2})}]_{\alpha'}=-i\epsilon_{2}^{a}\epsilon_{1}^{b}f_{ab}{}^{c}\quad\dot{X}_{c}(y_{1})e^{i(k_{1}+k_{2})\cdot X(y_{1})}+\text{scalars and tensors},
\end{equation}
where the structure constant $f_{ab}{}^{c}$ is given by (\ref{eq:structure-const}).

\subsection{Tensors and scalars contributions}

\label{sec:tensor-scalar}

We then proceed with the remaining tensor piece, $1$, and scalar parts,
$\epsilon_1\cdot \epsilon_2$ and  $(\epsilon_1\cdot
k_2)(\epsilon_2\cdot k_1)$, in (\ref{eq:bracket-v2v1}).
As was explained earlier, the equation should be understood to represent
only terms linear in all $\epsilon$'s, therefore when written explicitly
the factor $1$ term gives rise to a tensor operator
\begin{equation}
\int_{\tilde{\gamma}_{2}^{+}-\tilde{\gamma}_{2}^{-}}\frac{dy_{2}}{y_{2}}e^{\alpha'k_{1}\cdot k_{2}\ln(y_{1}-y_{2})}\left(:\epsilon_{1}\cdot\dot{X}(y_{1})e^{ik_{1}\cdot X(y_{1})}\epsilon_{2}\cdot\dot{X}(y_{2})e^{ik_{2}\cdot X(y_{2})}:\right)\,,\label{eq:tensor}
\end{equation}
whereas the $\epsilon_{1}\cdot\epsilon_{2}$ term corresponds to a
scalar
\begin{equation}
\int_{\tilde{\gamma}_{2}^{+}-\tilde{\gamma}_{2}^{-}}\frac{dy_{2}}{y_{2}}e^{\alpha'k_{1}\cdot k_{2}\ln(y_{1}-y_{2})}\frac{\epsilon_{2}\cdot\epsilon_{1}y_{1}y_{2}}{(y_{1}-y_{2})^{2}}\,\left(:e^{ik_{1}\cdot X(y_{1})}e^{ik_{2}\cdot X(y_{2})}:\right).\label{eq:scalar}
\end{equation}
As it happens, carrying out the integrals above can be a rather tricky
task. Taylor expanding (\ref{eq:tensor}) and (\ref{eq:scalar}) and
integrating the $y_{2}$ dependence term-wise leads to an infinite
sum of vertex operators associated with all possible levels. On the
other hand it is not clear whether such Taylor expansion would converge
in the first place, which can become an important issue for example
if we wish to incorporate more commutators and integrate further to
compute higher point numerators. For these reasons in the discussion
that follows we shall leave these two operator as they were and perform
the calculations directly on formulas (\ref{eq:tensor}) and (\ref{eq:scalar})
if needed.

As a double check, let us calculate the four-point $s$-channel numerator
$n(1234)$. In terms of commutators of vertex operators, this is given
by
\begin{equation}
n(1234)=\lim_{\alpha',y_{1},k_{4}^{2}\rightarrow0}\int_{\tilde{\gamma}_{3}^{+}-\tilde{\gamma}_{3}^{-}}\frac{dy_{3}}{y_{3}}\int_{\tilde{\gamma}_{2}^{+}-\tilde{\gamma}_{2}^{-}}\frac{dy_{2}}{y_{2}}\,\frac{1}{y_{1}}\Bigl\langle f\Bigr|[V_{3}(y_{3}),[V_{2}(y_{2}),V_{1}(y_{1})]_{\alpha'}]_{\alpha'}\Bigl|0\Bigr\rangle,\label{eq:string-4pt-numerator}
\end{equation}
where the final state is another level one vector (spin~1) particle, $\Bigl|f\Bigr\rangle=\epsilon_{4}\cdot\alpha_{-1}e^{ik_{4}\cdot x}\Bigl|0\Bigr\rangle$.
In this case we need not only the diffeomorphism part of the algebra,
but also the tensor and scalar described in (\ref{eq:tensor}) and
(\ref{eq:scalar}). To evaluate the $s$-channel numerator, in principle
we should include commutators of $V_{3}$ with these two operators
before sandwich them with a vacuum and final state.

It might help better organise the derivation if we know what to expect.
Recall that in \cite{Fu:2012uy,Fu:2016plh} the $s$-channel
numerator was shown to be
\begin{align}
n(1234)= & \ -\epsilon_{1}^{a}\epsilon_{2}^{b}\epsilon_{3}^{c}\epsilon_{4d}\,f_{ab}{}^{\sigma}f_{\sigma c}{}^{d}+\frac{1}{2}s_{12}\Bigl[(\epsilon_{1}\cdot\epsilon_{3})(\epsilon_{2}\cdot\epsilon_{4})-(\epsilon_{1}\cdot\epsilon_{4})(\epsilon_{2}\cdot\epsilon_{3})\Bigr]\nonumber \\
 & +\frac{1}{2}\Bigl(s_{23}-s_{13}\Bigr)(\epsilon_{1}\cdot\epsilon_{2})(\epsilon_{3}\cdot\epsilon_{4}) \,.\label{eq:ym-4pt-numerator}
\end{align}
We will see that the second term in (\ref{eq:ym-4pt-numerator}) can
be produced by commutator of $V_{3}$ and the tensor operator (\ref{eq:tensor}),
but has its value shifted; and similarly for the third term in the
equation, which corresponds to that of a scalar (\ref{eq:scalar}).
This difference between the two numerator prescriptions (\ref{eq:string-4pt-numerator})
and (\ref{eq:ym-4pt-numerator}) can be explained by the analytic
continuations introduced. We will see that the monodromy derived numerator
(\ref{eq:string-4pt-numerator}) reproduces the correct scattering
amplitude in the on-shell limit.

\subsection{Structure constants and disc integrals}

\label{sec:hyper-geometric-functions}

Consider first the operator product of $V_{3}$ with the tensor (\ref{eq:tensor}).
In order to produce an $\epsilon\cdot\epsilon$ term, operator $\epsilon_{3}\cdot\dot{X}$
has to Wick contract with either $\epsilon_{2}\cdot\dot{X}$ or $\epsilon_{1}\cdot\dot{X}$.
Both options actually lead to similar integrations. As a demonstration
of the derivation involved, let us focus on the $\epsilon_{2}\cdot\dot{X}$
contraction.
\begin{equation}
\int_{\tilde{\gamma}_{3}^{+}-\tilde{\gamma}_{3}^{-}}\frac{dy_{3}}{y_{3}}\int_{\tilde{\gamma}_{2}^{+}-\tilde{\gamma}_{2}^{-}}\frac{dy_{2}}{y_{2}}\,\frac{1}{y_{1}}\Bigl\langle f\Bigr|\Bigl(:\epsilon_{1}\cdot\dot{X}(y_{1})e^{ik_{1}\cdot X}\,\epsilon_{2}\cdot\underset{\text{contract}}{\dot{X}\underbracket{(y_{2})e^{ik_{2}\cdot X}:\Bigr)\,\epsilon_{3}\cdot\dot{X}}}(y_{3})e^{ik_{3}\cdot X}\Bigl|0\Bigr\rangle\,.\label{eq:e3e2-contraction}
\end{equation}
Standard Wick contraction replaces the contracted pair by the factor
$\alpha'y_{3}y_{2}(\epsilon_{3}\cdot\epsilon_{2})/(y_{3}-y_{2})^{2}$.
We then Taylor expand the remaining operators $\exp(ik_{3}\cdot X(y_{3}))$
and $\exp(ik_{2}\cdot X(y_{2}))$ around $y_{1}$ to extract all the
$y_{2}$ and $y_{3}$ dependence, in the hope that this manipulation
simplifies the integration. At lowest order we simply have a vector (spin~1)
vertex operator $\epsilon_{1}\cdot\dot{X}(y_{1})e^{i(k_{1}+k_{2}+k_{3})\cdot X(y_{1})}$,
so that in the $y_{1}\rightarrow0$ asymptotic limit this operator
becomes a level one state, $\lim_{y_{1}\rightarrow0}\,y_1^{-1}\,\epsilon_{1}\cdot\dot{X}(y_{1})e^{i(k_{1}+k_{2}+k_{3})\cdot X(y_{1})}\Bigl|0\Bigr\rangle=\epsilon_{1}\cdot\alpha_{-1}e^{-ik_{4}\cdot x}\Bigl|0\Bigr\rangle$,
which subsequently contracts with the final state, yielding a factor
$(\epsilon_{1}\cdot\epsilon_{4})$. All higher power terms in the
Taylor expansion can be argued to produce different asymptotic states,
thereby giving no contribution to the expectation value\footnote{Alternatively, higher power terms can be argued to contain higher
powers of $\alpha'$ as well, therefore not contributing. This is
because the $X(y)$ operator carries an overall $\sqrt{\alpha'}$
factor. }. In other words, equation (\ref{eq:e3e2-contraction}) yields
\begin{align}
 & \int_{\tilde{\gamma}_{3}^{+}-\tilde{\gamma}_{3}^{-}}dy_{3}\int_{\tilde{\gamma}_{2}^{+}-\tilde{\gamma}_{2}^{-}}dy_{2}\,e^{\alpha'k_{2}\cdot k_{3}\ln(y_{2}-y_{3})}e^{\alpha'k_{1}\cdot k_{3}\ln(y_{1}-y_{3})}e^{\alpha'k_{1}\cdot k_{2}\ln(y_{1}-y_{2})}\nonumber \\
 & \hspace{5cm} \frac{\epsilon_{2}\cdot\epsilon_{3}}{(y_{2}-y_{3})^{2}}\Bigl\langle f\Bigr|:\epsilon_{1}\cdot\dot{X}(y_{1})e^{ik_{1}\cdot X(y_{1})}e^{ik_{2}\cdot X(y_{2})}e^{ik_{3}\cdot X(y_{3})}:\Bigl|0\Bigr\rangle\nonumber \\
 & =(\epsilon_{2}\cdot\epsilon_{3})(\epsilon_{1}\cdot\epsilon_{4})\int_{\tilde{\gamma}_{3}^{+}-\tilde{\gamma}_{3}^{-}}dy_{3}\int_{\tilde{\gamma}_{2}^{+}-\tilde{\gamma}_{2}^{-}}dy_{2}e^{\alpha'k_{2}\cdot k_{3}\ln(y_{2}-y_{3})}e^{\alpha'k_{1}\cdot k_{3}\ln(y_{1}-y_{3})}e^{\alpha'k_{1}\cdot k_{2}\ln(y_{1}-y_{2})}\frac{1}{(y_{2}-y_{3})^{2}}
\end{align}
Our remaining task is then to evaluate the above integral. At first
sight a direct computation can be slightly difficult. 
However we can always resort to the derivation used  to
extract momentum kernel from monodromy relations \cite{BjerrumBohr:2010hn}
and apply the same reasoning to translate the double integral to something
recognisable. Explicitly we write
\begin{align}
 & \int_{\tilde{\gamma}_{3}^{+}-\tilde{\gamma}_{3}^{-}}dy_{3}\int_{\tilde{\gamma}_{2}^{+}-\tilde{\gamma}_{2}^{-}}dy_{2}e^{\alpha'k_{2}\cdot k_{3}\ln(y_{2}-y_{3})}e^{\alpha'k_{1}\cdot k_{3}\ln(y_{1}-y_{3})}e^{\alpha'k_{1}\cdot k_{2}\ln(y_{1}-y_{2})}\frac{1}{(y_{2}-y_{3})^{2}}\nonumber \\
 & =\sin(\pi\alpha'k_{2}\cdot k_{1})\sin(\pi\alpha'k_{3}\cdot k_{1})\,I(1324)+\sin(\pi\alpha'k_{2}\cdot k_{1})\sin(\pi\alpha'(k_{3}\cdot k_{1}+k_{3}\cdot k_{2}))\,I(1234),\label{eq:e3e2-eqn2}
\end{align}
where
\begin{equation}
I(1234)=\int_{0}^{1}dy_{3}\int_{0}^{y_{3}}dy_{2}e^{\alpha'k_{3}\cdot k_{2}\ln(y_{3}-y_{2})}e^{\alpha'k_{3}\cdot k_{1}\ln(y_{3}-y_{1})}e^{\alpha'k_{2}\cdot k_{1}\ln(y_{2}-y_{1})}\frac{1}{(y_{3}-y_{2})^{2}},
\end{equation}
and
\begin{equation}
I(1324)=\int_{0}^{1}dy_{2}\int_{0}^{y_{2}}dy_{3}e^{\alpha'k_{3}\cdot k_{2}\ln(y_{3}-y_{2})}e^{\alpha'k_{3}\cdot k_{1}\ln(y_{3}-y_{1})}e^{\alpha'k_{2}\cdot k_{1}\ln(y_{2}-y_{1})}\frac{1}{(y_{3}-y_{2})^{2}}.
\end{equation}
From a simple change of integration variables we see that both $I(1234)$
and $I(1324)$ are actually proportional to beta functions.
\begin{align}
I(1234)  = & \int_{0}^{1}dy_{3}\frac{1}{y_{3}^{2}}e^{\alpha'(k_{3}\cdot k_{2}+k_{3}\cdot k_{1}+k_{2}\cdot k_{1})\ln\,y_{3}}\int_{y_{2}=0}^{y_{2}=y_{3}}y_{3}d\left(\frac{y_{2}}{y_{3}}\right)\nonumber \\
 & \frac{1}{\left(1-\left(\frac{y_{2}}{y_{3}}\right)\right)^{2}}e^{\alpha'k_{3}\cdot k_{2}\ln\left(1-\left(\frac{y_{2}}{y_{3}}\right)\right)}e^{\alpha'k_{3}\cdot k_{1}\ln\left(1-\left(\frac{y_{1}}{y_{3}}\right)\right)}e^{\alpha'k_{2}\cdot k_{1}ln\left(\left(\frac{y_{2}}{y_{3}}\right)-\left(\frac{y_{1}}{y_{3}}\right)\right)}\,.
\end{align}
Suppose if we introduce a new variable $y_{2}'=(y_{2}/y_{3})$ and
let $y_{1}\rightarrow0$, this becomes
\begin{align}
 & \int_{0}^{1}dy_{3}\,y_{3}^{\alpha'(k_{3}\cdot k_{2}+k_{3}\cdot k_{1}+k_{2}\cdot k_{1})-1}\int_{0}^{1}dy_{2}'(1-y_{2}')^{\alpha'k_{3}\cdot k_{2}-2}y_{2}^{\alpha'k_{2}\cdot k_{1}}\nonumber \\
 & =\frac{1}{\alpha'k_{4}^{2}}\,B(\alpha'k_{3}\cdot k_{2}-1,\,\alpha'k_{2}\cdot k_{1}+1)
\end{align}
where we used momentum conservation $k_{3}\cdot k_{2}+k_{3}\cdot k_{1}+k_{2}\cdot k_{1}=k_{4}^{2}$
and the integral definition of beta function. 
This expression is an explicit realisation of the  $\alpha'$-dependent
Berends-Giele currents  $\phi_{ A | B }$  of~\cite{Mafra:2016mcc}.

In the $\alpha'\rightarrow0$ limit the beta function yields
\begin{align}
\lim_{\alpha'\to0}B(\alpha'k_{3}\cdot k_{2}-1,\,\alpha'k_{2}\cdot k_{1}+1) & =\lim_{\alpha'\to0}\frac{\Gamma(\alpha'k_{3}\cdot k_{2}-1)\,\Gamma(\alpha'k_{2}\cdot k_{1}+1)}{\Gamma(\alpha'k_{3}\cdot k_{2}+\alpha'k_{2}\cdot k_{1})}\nonumber \\
 & =(-1)\frac{k_{3}\cdot k_{2}+k_{2}\cdot k_{1}}{k_{3}\cdot k_{2}}\,.
\end{align}
Repeating similar derivation on $I(1324)$ and plugging everything
back shows that equation (\ref{eq:e3e2-eqn2}) gives
\begin{align}
 & (\epsilon_{3}\cdot\epsilon_{2})(\epsilon_{1}\cdot\epsilon_{4})\frac{s_{21}}{k_{4}^{2}}\left[s_{31}(-1)\frac{s_{13}+s_{23}}{s_{23}}+(s_{31}+s_{32})(-1)\frac{s_{23}+s_{12}}{s_{23}}\right]\nonumber \\
 & =(\epsilon_{3}\cdot\epsilon_{2})(\epsilon_{1}\cdot\epsilon_{4})(-1)s_{12}\frac{s_{13}+s_{23}}{s_{23}}.
\end{align}

Generically an overall rescaling produces a $1/k_{n}^{2}$ as in the
four-point example. The remaining integral can be identified with the
$Z$-function  in~\cite{Mafra:2016mcc} when appropriately integrated by
parts.

\subsection{Jacobi identities and the four-point numerators}

\label{sec:jacobi-4pt}

In the previous section we demonstrated that the $s$-channel numerator
$n(1234)$ can be calculated from the expectation value of successive
$\alpha'$-weighted commutators of vertex operators $[[V_{1},V_{2}]_{\alpha'},V_{3}]_{\alpha'}$.
It is straightforward to see that the $u$-channel $n(1324)$ can
be obtained by the same reasoning, and the result is that of the $s$-channel
with legs $2$ and $3$ swapped. Note however, that the $t$-channel
numerator on the other hand is not related to the other two by a simple
relabelling. This is because the asymptotic state condition requires
that the world-sheet coordinate $y_{1}\rightarrow0$ rather than being
integrated, so that in our settings leg $1$ and the rest of the legs
were not placed on equal footing. Instead, the $t$-channel numerator
$\Bigl\langle4\Bigr|[[V_{1},V_{2}]_{\alpha'},V_{3}]_{\alpha'}\Bigl|0\Bigr\rangle$
requires evaluating the following integrals.
\begin{align}
 & \int\prod_{i=1}^{3}\frac{dy_{i}}{y_{i}}V_{1}(y_{1})V_{2}(y_{2})V_{3}(y_{3})-e^{-i\pi\alpha'k_{1}\cdot k_{2}}\int\prod_{i=1}^{3}\frac{dy_{i}}{y_{i}}V_{2}(y_{2})V_{1}(y_{1})V_{3}(y_{3})\nonumber \\
 & -e^{-i\pi\alpha'k_{3}\cdot(k_{1}+k_{2})}\int\prod_{i=1}^{3}\frac{dy_{i}}{y_{i}}V_{3}(y_{3})V_{1}(y_{1})V_{2}(y_{2})+e^{-i\pi\alpha'(k_{1}\cdot k_{2}+k_{2}\cdot k_{3}+k_{3}\cdot k_{1})}\int\prod_{i=1}^{3}\frac{dy_{i}}{y_{i}}V_{3}(y_{3})V_{2}(y_{2})V_{1}(y_{1}),
\end{align}
where the integration contour of each term follows the convention
described in section \ref{sec:the-algebra}, and can be read off directly
from the order of the original vertex operators. To carry out the
integral above, we note that from the definition of the $\alpha'$-weighted
commutator, we have
\begin{align}
 & [[V_{1},V_{2}]_{\alpha'},V_{3}]_{\alpha'}-[[V_{1},V_{3}]_{\alpha'},V_{3}]_{\alpha'}\nonumber \\
 & =[V_{1},[V_{2},V_{3}]_{\alpha'}]_{\alpha'}+(e^{-i\pi\alpha'k_{2}\cdot k_{3}}-1)\Biggl(V_{1}V_{3}V_{2}+e^{-i\pi\alpha'k_{1}\cdot k_{2}}V_{2}V_{1}V_{3}-e^{-i\pi\alpha'k_{1}\cdot k_{3}}V_{3}V_{1}V_{2}-e^{-i\pi\alpha'k_{1}\cdot(k_{2}+k_{3})}V_{2}V_{3}V_{1}\Biggr).
\end{align}
The second term above carries a factor $(e^{-i\pi\alpha'k_{2}\cdot k_{3}}-1)\sim\mathcal{O}(\alpha')$
so that it vanishes in the small $\alpha'$ limit, and therefore $n_{s}+n_{t}+n_{u}=0$.

Summarising the results, at four points we have 
\begin{align}
n(1234) & =n_{s}=\lim_{\alpha',y_{1},k_{4}^{2}\rightarrow0}\int_{\tilde{\gamma}_{3}^{+}-\tilde{\gamma}_{3}^{-}}\frac{dy_{3}}{y_{3}}\int_{\tilde{\gamma}_{2}^{+}-\tilde{\gamma}_{2}^{-}}\frac{dy_{2}}{y_{2}}\,\frac{1}{y_{1}}\Bigl\langle f\Bigr|[[V_{1}(y_{1}),V_{2}(y_{2})]_{\alpha'},V_{3}(y_{3})]_{\alpha'}\Bigl|0\Bigr\rangle\nonumber \\
 & =\epsilon_{1}^{a}\epsilon_{2}^{b}\epsilon_{3}^{c}\epsilon_{4d}\,f_{ab}{}^{\sigma}f_{\sigma c}{}^{d}+s_{23}(\epsilon_{1}\cdot\epsilon_{2})(\epsilon_{3}\cdot\epsilon_{4})-s_{12}\frac{s_{13}+s_{23}}{s_{23}}(\epsilon_{3}\cdot\epsilon_{2})(\epsilon_{1}\cdot\epsilon_{4})\\
n(1324) & =-n_{u}=\lim_{\alpha',y_{1},k_{4}^{2}\rightarrow0}\int_{\tilde{\gamma}_{2}^{+}-\tilde{\gamma}_{2}^{-}}\frac{dy_{2}}{y_{2}}\int_{\tilde{\gamma}_{3}^{+}-\tilde{\gamma}_{3}^{-}}\frac{dy_{3}}{y_{3}}\,\frac{1}{y_{1}}\Bigl\langle f\Bigr|[[V_{1}(y_{1}),V_{3}(y_{3})]_{\alpha'},V_{2}(y_{2})]_{\alpha'}\Bigl|0\Bigr\rangle\nonumber \\
 & =\epsilon_{1}^{a}\epsilon_{2}^{b}\epsilon_{3}^{c}\epsilon_{4d}\,f_{ac}{}^{\sigma}f_{\sigma b}{}^{d}+s_{23}(\epsilon_{1}\cdot\epsilon_{3})(\epsilon_{2}\cdot\epsilon_{4})-s_{13}\frac{s_{12}+s_{23}}{s_{23}}(\epsilon_{3}\cdot\epsilon_{2})(\epsilon_{1}\cdot\epsilon_{4})\\
n_{t} & =n(1234)-n(1324)\nonumber \\
 & =\epsilon_{1}^{a}\epsilon_{2}^{b}\epsilon_{3}^{c}\epsilon_{4d}\,f_{bc}{}^{\sigma}f_{\sigma a}{}^{d}-s_{23}(\epsilon_{1}\cdot\epsilon_{2})(\epsilon_{3}\cdot\epsilon_{4})+s_{23}(\epsilon_{1}\cdot\epsilon_{3})(\epsilon_{2}\cdot\epsilon_{4})+(s_{12}-s_{13})(\epsilon_{3}\cdot\epsilon_{2})(\epsilon_{1}\cdot\epsilon_{4})
\end{align}
A straightforward calculation shows that the above set of numerators
does reproduce the correct four-point Yang-Mills amplitudes.


\section{Conclusion}
\label{sec:conclusion}

In this paper, we have presented a vertex operator algebra construction
of the BCJ kinematic numerators. Our construction provides an
alternative construction to the kinematic traces
of~\cite{Bern:2011ia}. Starting with the introduction of a formal off-shell
continuation to the open string amplitude, we implemented the analytic procedure
used to derive the string momentum kernel in \cite{BjerrumBohr:2010hn}, and
obtained an integral formula that can be taken as a string analogue
of the kinematic numerator. We next translated the integral form into
the vertex operator language, and the kinematic numerator naturally
gave rise to the structure of successive $\alpha'$-weighted commutators.
This result mirrors the colour structure of open string and semi-Abelian
$Z$-theory \cite{Ma:2011um,Carrasco:2016ygv} obtained by employing monodromy
relations in the colour decomposition formulas. As a comparison to
previous discussions on algebraic interpretation of the Yang-Mill
numerator, we reproduced the algebra of diffeomorphism as the vector
$\times$ vector $\rightarrow$ vector part of the vertex operator
algebra. We showed at four points the numerator is restored when the
full algebra is taken into account. We showed in general, that the
string of structure constants is given by hypergeometric functions
from disc integrals.

Our newly obtained string analogue numerator, when plugged into the
KLT monodromy relation produces a string version of the BCJ dual Del
Duca-Dixon-Maltoni formula. Comparing the results with
\cite{Ma:2011um,Carrasco:2016ygv} the present construction suggests that one can obtain
string generalisation of many double-copied field theory
amplitudes by replacing all the commutators by the
$\alpha'$-weighted and the field theory amplitudes with the
corresponding open string ones, therefore obtaining new classes of string theory
tree-level amplitudes. It is reasonable to expect that some Feynman
diagram-like description of the open string amplitude exists if we
follow this line of thoughts and exploit the $\alpha'$-weighted
commutators or its generalisations. It would be interesting to see if
the double copy expression also came from a string origin, which might
shed light on loop level structures. As well, it would be interesting
to understand if string theory provides additional constraints on
these new classes of tree-level amplitudes.

From the vertex operator viewpoint it is not clear to us why the kinematic
numerator derived in this paper corresponds to field theory amplitudes
(namely Yang-Mills) given by Feynman rules that are limited to three
and four point interactions only. It is also not entirely clear whether
the cubic graph organisation leads to the same decomposing of the
Yang-Mills quartic vertex as the recent prescription developed in the
context of CHY formulation \cite{Bjerrum-Bohr:2016axv,Huang:2017ydz,Du:2017kpo}. We leave these interesting
questions to future work.

\section*{Acknowledgements}

We would like to thank N.E.J. Bjerrum-Bohr, Donal O'Connell, Yi-Jian Du, Bo Feng, Song
He, Kirill Krasnov and Oliver Schlotterer, for
valuable discussions and comments on the manuscript. CF is particularly
grateful for Kirill Krasnov for suggesting this project. The research
of CF is supported by the Fundamental Research Funds for the Central
Universities (GK201803018). P. Vanhove has received funding the ANR grant ``Amplitudes'' ANR-17-
CE31-0001-01, and is partially supported by Laboratory of Mirror
Symmetry NRU HSE, RF Government grant, ag. N$^\circ$ 14.641.31.0001.
Yihong Wang is supported by MoST grant 106-2811-M-002-196.
Both CF and YW are thankful for the Institute of 
Theoretical Physics, CAS in Beijing for hospitality during this work. 




\end{document}